\documentclass{svjour3}
\usepackage[american]{babel}
\usepackage{graphicx,csquotes,xcolor,enumerate,amsmath,amssymb,mathtools,amsfonts,epigraph,setspace,indentfirst}
\usepackage{newtxtext,newtxmath}

\usepackage[margin=3cm]{geometry}

\journalname{Forthcoming in \textit{Foundations of Physics}}

\title{Antirealism in sheep's clothing}
\subtitle{Wave function realism and scientific realism}
\author{Raoni Arroyo \and Jonas R. Becker Arenhart}

\institute{Raoni Arroyo \at Nucleus of Epistemology and Logic (NEL), Federal University of Santa Catarina (UFSC), Florianópolis, Brazil \at Research Group on Logic and Foundations of Science (CNPq)\at Corresponding author: \email{raoniarroyo@gmail.com} \and Jonas R. Becker Arenhart \at Department of Philosophy, Federal University of Santa Catarina, Florianópolis, Brazil \at Research Group on Logic and Foundations of Science (CNPq) \at \email{jonas.becker2@gmail.com}}

\date{\today}

\begin{document}
\sloppy\raggedbottom\maketitle

\begin{abstract}
Scientific realism is the philosophical stance that science tracks truth, in particular in its depiction of the world's ontology. Ontologically, this involves a commitment to the existence of entities posited by our best scientific theories; metaontologically, it includes the claim that the theoretical framework itself is true. In this article, we examine wave function realism as a case study within this broader methodological debate. Wave function realism holds that the wave function, as described by quantum mechanics, corresponds to a real physical entity. We focus on a recent formulation of this view that commits to the ontology of the wave function while deliberately avoiding the metaontological question of the framework's truth. Instead, the view is defended on pragmatic, non-truth-conductive grounds. This, we argue, raises tensions for the purported realism of wave function realism and its compatibility with scientific realism more broadly.

\end{abstract}

\epigraph{``I am sorry that we can never quite reach rock bottom in the vindication of a framework and that we cannot stand back in metaphysical grandeur and make our philosophical decisions independently of all frameworks, but, as Bertrand Russell says, it is not my fault.''}{Grover Maxwell \cite[137]{maxwell1962}.}

\section{Introduction}

In a very broad way, scientific realism claims that scientific theories are approximately true, so we can learn how the world is if we pay close attention to them. Because of this alethic requirement, an ontological consequence is straightforward: we should believe in the existence of the entities posited by our best scientific theories, as they are true descriptions of the world.

Wave function realism (WFR) seems to check all the boxes for scientific realism. Roughly speaking, WFR is the view according to which wave functions, as described by quantum mechanics, are real entities. In other words, the world is populated by wave functions that are not only part of the representational apparatus of the theory. The hope of WFR, as may be expected, is that quantum mechanics itself warrants this claim. 

This makes WFR the perfect example to build a general case on how realism (about science or about wave functions) should operate: realists should hold that the ontological stuff they claim realism about should be the ontological stuff really---truly!---there in the world. We call this the metaontological question. In other words, the metaontological question states that there is a correspondence between what the theory says and how the world really is.

In order to get there, this paper discusses a specific approach to WFR: an approach that we'll call ``high-dimensional wave function realism'' hereafter ``WFR$^{HD}$''. This approach is notoriously put forth and defended by Alyssa Ney. To the unfamiliar reader on WFR, a warning is in order. WFR is not a unified position, being elaborated by distinct philosophers. As such, one can find some distinct variants of the view in the literature (for an updated survey, see \cite[sect.~3]{chen2019}). For instance, Ney's and David Albert's approaches are called ``\textit{ontological}''---because both of them consider ``quantum mechanical wave functions \textelp{} as concrete physical objects'' \cite[53]{albert2013}---but even these versions differ substantially. It is beyond the scope of this paper, however, to map such differences (and the interested reader will find such a map in \cite{chen2019}). This paper focuses exclusively on WFR$^{HD}$, which is, to our best knowledge, the most fine-grained ontological proposal of WFR available (and discussed) in the literature---see, for instance, \cite{wallace2021routledge,hubert-2022,maudlin2022}. 

After presenting what we take the scientific realism debate to be, viz., the metaontological question (\S \ref{sec:2}), we analyze how WFR$^{HD}$ is defended by Ney on Carnapian grounds (\S \ref{sec:3}) and conclude it's not up to realist standards (\S \ref{sec:4}). In calibrating its goals according to what it actually delivers, however, the notion of realism seems to give room to its antirealist oponents. WFR$^{HD}$ is particularly telling on this, as there are other genuinely---metaontologically---realist pictures of WFR (\S \ref{sec:5}). So, our argument takes a specific strain of wave function realism as its case study, but we contend that its implications can be extended to scientific realism more generally.

\section{Ontology, metaontology, and realism}\label{sec:2}

Let's start with realism. There are many characterizations of what scientific realism is, and we agree with \cite[sect.~1.1]{chakravartty-sep-scientific-realism} that ``[i]t is perhaps only a slight exaggeration to say that scientific realism is characterized differently by every author who discusses it''. That said, there are two closely related common denominators to all of them.\footnote{See also Anjan Chakravartty: ``Amidst these differences, however, a general recipe for realism is widely shared: our best scientific theories give true or approximately true descriptions of observable and unobservable aspects of a mind-independent world'' \cite{chakravartty-sep-scientific-realism}.} The first one is the claim that scientific theories are about the \textit{world} \cite[150]{fine1986}. This is cashed out in terms of science telling us: ``how the world is'' (\cite[324]{french2018}; \cite[369]{kincaid2018}); ``what exists in the world'' (\cite[104]{maudlin2007}; \cite[419]{bird2018}); ``true descriptions of the world'' (\cite[19]{hoefer2020}; \cite[57]{callender2020};  \cite[18]{mizrahi2020}), sometimes explicitly fleshed out as a correspondentist account of truth (\cite[96]{ladymanross2007}; \cite[293]{ruetsche2020}). The second one is the notion of `belief' \cite{smart1963}, \cite{boyd1983}, \cite{devitt1991}, \cite{kukla1998}, \cite{niiniluoto1999}, \cite{psillos1999}, \cite{chakravartty2007}. If there are aspects of scientific theories that truly latch onto the world (e.g., the ontological posits of these theories) in a correspontential sense, then we are recommended to believe that such entities exist. Emphatically enough, here's how Arthur Fine draws the realist picture we have in mind:

\begin{quote}
    For realism, science is \textit{about} something; something \textit{out there}, `external' and (largely) independent of us. The traditional conjunction of externality and independence leads to the realist picture of an objective, external world; what I shall call the \textit{World}. According to realism, science is \textit{about that}. Being about the \textit{World} is what gives significance to science. That is, on the semantic side, we are to understand scientific claims as claims about the \textit{World}. Thus, realism adopts a special interpretative stance towards the language of science, the stance traditionally assumed by a correspondence theory of truth and referential semantics, provided the referents are (in general) taken as Real; i.e. as elements of the \textit{World}. \cite[150]{fine1986}
\end{quote}

There are cases in which the realist perspective isn't necessarily advocated for the entire theory's content, it applies to certain aspects (e.g., causally effective entities, structures, dispositions, etc.). This is often called ``partial'' realism. It should be noted that even in these cases, the realist stance involves reifying these elements in the world, precisely \textit{because} they are (at least partially) true.

These matters could be framed into the Carnapian well-documented internal-external divide. For Rudolf Carnap, all meaningful theories require the formulation of a corresponding linguistic framework, which contains (i) expressions from the language of that framework, along with formation rules, and (ii) a set of rules for evaluating whether an expression is true or false in the framework \cite{carnap1950}. For example, an expression such as ``the squared norm of the wave function is the probability distribution over an observable operator'' would be a linguistic expression of standard formulations of quantum mechanics. And, as there is a linguistic system with specific rules---a \textit{framework}---that guarantees that we can connect experimental results with this expression, it can be considered as supported in this framework. According to the standard way of thinking about this, only the ontological/internal issues are meaningful. By borrowing a little bit of terminology from David Chalmers, then, we have the following `internal \textit{qua ontological}' and `external \textit{qua metaontological}' divide:

\begin{quote}
The basic question of ontology is ``What exists?''. The basic question of metaontology is: are there objective answers to the basic question of ontology? \cite[77]{chalmers2009}
\end{quote}

In this way, scientific realism should address in some sense the metaontological question. Otherwise, it cannot meaningfully talk about \textit{truth} and the \textit{world} and, consequently, ask for \textit{belief} in its ontological content---which, as we saw, are among its defining aspects.

Consider an ontological question, such as: \textit{do wave functions exist?} This issue can be internal or external with respect to a framework. If it were an internal question, it would be ``quickly decidable'' \cite[132]{maxwell1962}: wave functions do exist within certain frameworks of quantum mechanics---after all, they are part of (i), that is, of the expressions postulated by the language of certain frameworks. The time-dependent standard Schrödinger equation, for instance, describes the evolution of the \textit{wave function} in time.

This is a familiar issue in the Carnapian philosophy of science. So let's apply such a terminology to WFR$^{HD}$. Following a closely related discussion raised by James Ladyman \cite[154]{ladyman2010manyworlds} concerning many worlds and quantum mechanics, there seems to be three basic questions that must be distinguished when discussing WFR$^{HD}$:

\begin{enumerate}
    \item Is WFR$^{HD}$ coherent?
    \item Is WFR$^{HD}$ the best ontology of quantum mechanics?
    \item Are there wave functions?
\end{enumerate}

We'll take the first question for granted, as WFR$^{HD}$ is already a well-established view in the literature \cite{ney-albert2013wfr}. Our focus is on a specific action taken by WFR$^{HD}$, which involves responding affirmatively to the second question by utilizing the affirmative answer to the third question. Both align in the internal-external divide. The internal/ontological question is: ``are there wave functions?'', to which one might answer ``yes'' or ``no'', depending on the formulation of quantum mechanics one adopts; whereas the external/metaontological question is ``is there an objective fact of the matter about these answers?''. WFR$^{HD}$, as a realist view on quantum ontology, should answer affirmatively to the metaontological question.

\section{Ontological (wave function) realism}\label{sec:3}

Most discussions on quantum ontology begin with solutions to the measurement problem(s). The odds of this one being different are low, since ``\textelp{} the meaning of the wave function is related to solutions to the quantum measurement problem'' \cite[e12612]{chen2019}. But not every discussion is on the meaning of the wave function. As advanced by Ney, the wave function is a commonality between \textit{all} known solutions to the measurement problem: ``\textelp{} each interpretation of the quantum formalism makes central use of a wave function \textelp{}'' \cite[34]{ney2021oup}. With this claim taken at face value, WFR$^{HD}$ has been proposed as a natural way of \textit{reading off} ontological commitments of quantum mechanics: because quantum mechanics cannot dispense with wave functions, the latter must be taken ontologically seriously, i.e., there are wave functions.

This is the ``prima facie'' argument for WFR$^{HD}$ \cite{ney2012philstud}: ontology should be continuous with science; wave functions are indispensable for quantum mechanics; therefore, we should assume that wave functions are real entities described by quantum mechanics. In other words, we should adopt a realist stance toward the wave function. Notice that the argument expects to go from the indispensability of a mathematical apparatus to the formulation of a theory to the commitment with a physical counterpart to that mathematical entity. That is more than typical indispensability arguments do in mathematics: what is expected to be achieved by such arguments is commitment with mathematicalia, not with any additional physical counterparts they may have. In the case of WFR, the argument expects to extract physical commitments from a mathematical indispensability. There are two problems with that, all related to underdetermination.

On the one hand, the slide from the mathematical apparatus to the physical counterpart may be resisted by traditional strategies such as quantifying over mathematical entities while refraining to believe that they represent any physical entities \cite{maudlin2013}. Additionally, one may use neutral quantifiers and avoid the indispensability argument altogether \cite{bueno2005}. But that is not the most radical way to stop the use of the indispensability argument in this case. There are even more radical challenges advanced against it by directly attacking its major premise, viz., the one stating that wave functions are indispensable to the formulation of quantum mechanics. Critical discussions on WFR highlighted this fact by dropping the indispensability claim of wave functions; that was achieved with the advancement of many formulations of quantum theory \emph{not even mentioning wave functions} \cite{bokulich2020wfr}, and today there is a consensus that an ontology of wave functions is not read off from the formalism of every formulation of the theory \cite{wallace2021routledge}. The obvious way to resist this argument would be to suggest that different formulations amount to different theories, but doing that, we presume, would be laboring on a very unorthodox approach to theory identity, even for the standards of quantum mechanics. 

Even though the prima facie argument isn't how WFR$^{HD}$ is defended nowadays \cite[4]{ney2023}, and despite reasons for resisting indispensability, we may concede the following: there are wave functions at least within specific formulations of quantum mechanics. How does that leave us on the issue of realism? We still need to grant that such formulations requiring wave functions are approximately true. That is, we need to go from ``there are formulations of quantum mechanics where wave functions exist'' to the stronger claim that such formulations are true, or approximately true. To do so, an extra ingredient needs to be called in: one needs to answer the metaontological basic question.\footnote{Discussing WFR$^{HD}$, Ney explicitly claims that we need to be a realist about quantum mechanics, after all: ``Of course, we should be realists about quantum theories. Quantum theories, like the quantum field theories that make up the Standard Model of particle physics, have been extremely well-confirmed. And we should be realists about quite a lot of the phenomena quantum theories reveal'' \cite[15]{ney2023}. More on this topic later.} 

\section{Metaontological (wave function) empiricism}\label{sec:4}

On (what we are calling) metaontological grounds, WFR$^{HD}$ is defended employing a Carnapian strategy, and this comes along with a price to pay. Recall that, when it comes to the ontology of our theories, we have reason to believe that the posits of such theories \emph{exist from the point of view of the theory itself}. Thus, there \textit{is} an objective fact of the matter about the ontology---the existence of such and such entities---as \textit{internal} to a given framework. Here is where realism scores: from an internal-question point of view, one may point out framework-relative objective reasons for believing the answer to the ontological question (viz., are there wave functions in such-and-such formulation of quantum mechanics?) is ``yes''.

Answers to the metaontological question are not that straightforward. Is the ontological framework of WFR$^{HD}$ true? We don't know. We just don't have the resources to grant that a specific framework (and its corresponding ontological aspects) is approximately true. Which is to say that we don't have the required resources to grant an objectively true answer to the metaontological question for the quantum domain. 

Physics stops, at most, at the ontological level. That is, while commitments inside a framework may be justified by the framework's rules, the same cannot be said for the additional claim to the truth of the framework itself. In these terms, a realist wants to know whether wave functions exist in the world---and here we are raising questions  that are meaningless outside a linguistic framework. There is a gap between ontology and metaontology (which concerns itself with the epistemological questions for the whole framework), and we believe  scientific realist are the ones interested in answering the metaontological question. So, are there wave functions? Within WFR$^{HD}$, yes (the ontological realism); but are there wave functions out there in the world (metaontological realism)? This we cannot answer, partly because we don't yet have the right tools---meaning: physics---to objectively decide which is the framework whose entities truly correspond to the world.

There's no knockdown argument favoring WFR, nor WFR$^{HD}$, and everyone is aware of that. This is partly why Ney adopts a Carnapian stance of ``\textelp{} humility and tolerance for other approaches'' \cite[XI]{ney2021oup}.

Yet, WFR$^{HD}$ is defended on pragmatic grounds by the \textit{argument from entanglement}, put forth in \cite{ney2012nous} and \cite[cha2]{ney2021oup}, which briefly runs as follows. Without understanding wave functions as real high-dimensional entities, one cannot distinguish in one's ontology between empirically distinct physical states, e.g. separable and non-separable (i.e. entangled) states---which can be done in a high-dimensional wave-function ontology. 

Of course, the argument from entanglement does not entail a straightforward case for the metaontological question. Again, this is so because WFR (\textit{qua} WFR$^{HD}$ or not) is not the only game in town. Other approaches do the same job without positing wave-function or high-dimensional (or both!) ontologies.\footnote{A rich survey of them can be found in \cite[56--76]{ney2021oup}, but only to give a short sampling we may mention, e.g.: low-dimensional holism \cite{howard1989,teller1986}; multi-field approach \cite{hubert-romano-2018}; ontic structural realism \cite{french2014,frenchladyman2003}; primitive ontology approach \cite{allori2013}; spacetime state realism \cite{wallace-timpson2010}; the property view \cite{monton2013}.} As might be expected, the metaontological question concerning the ontology of quantum mechanics that truly corresponds to the world is underdetermined by its alternatives. And this is also acknowledged by WFR$^{HD}$; here is Ney:

\begin{quote}
    The very existence and empirical adequacy of these alternatives undermines the argument from entanglement, demonstrating that it is not true that wave function realism is essential for providing a realist characterization of quantum mechanics, including entangled states. \cite[56]{ney2021oup}.
\end{quote}

So maybe one can find another road. Here's \textit{also} Ney ``\textelp{} even if the facts of entanglement underdetermine an answer to the ontological question, this does not mean a case cannot be made for preferring WFR$^{HD}$ over rival frameworks'' \cite[81]{ney2021oup}. As far as it goes, it seems that WFR$^{HD}$ is a proposal on the metaontological question. Such a proposal, however, is not defended as the true one. Because of its rival underdetermined options, endorsing WFR$^{HD}$ is not a matter of truth, but of pragmatism. It's defended by being \textit{preferable}, not by being true. This generates the third argument for WFR$^{HD}$: the ``argument from locality and separability'': 

\begin{quote}
    \textelp{} wave function realism [qua WFR$^{HD}$] is unique in yielding quantum ontologies that not only distinguish quantum states, but do so by retaining two intuitively nice metaphysical features: separability and locality. \cite[8]{ney2023}.
\end{quote}

\begin{quote}
    \textelp{} having separable and local interpretations of our best physical theories yields both theoretical and practical benefits, this is a reason to favor wave function realism as an approach to interpreting quantum theories. \cite[81]{ney2021oup}.
\end{quote}

\begin{quote}
    It is my view that the best case the wave function realist has for developing her distinctive ontological framework comes from such conceptual considerations and intuitions. \textelp{} The discovery and development of interpretations that are compatible with our intuitions may be useful for a number of reasons. \textelp{} All are unabashedly pragmatic. \cite[129--130]{ney2021oup}.
\end{quote}

Indeed, arguably the most counterintuitive features of quantum mechanics are non-locality and non-separability---and this can be either a feature or a bug, depending on how one couches the evidential role of intuitions \cite{mizrahi2022}. Is this compatible with scientific realism? Here's how Ney understands the term in her project of wave function \textit{realism}:

\begin{quote}
    Scientific realists are those who believe that our best scientific theories are \textit{true (or approximately true)} and that the entities they describe exist in at least roughly the ways these theories say they do. \cite[69, emphasis added]{ney2021oup}.
\end{quote}

WFR$^{HD}$ is thus the view according to which ``\textelp{} wave functions are not just real, but are real, objective, physical fields'' \cite[69]{ney2021oup}. The defense for this characterization, however, is not cashed out in metaontological/external terms of truth and of \textit{objective} existence. It is rather made in terms of pragmatic considerations and internal/ontological, framework-dependent existence. So that characterization isn't properly justified. That's a consequence of playing the game with the Carnapian rules: the metaontological question will never be answered either in terms of truth or in terms of correspondence of the world---it might be a pragmatic question of adopting the WFR$^{HD}$ framework \textit{at best}. That's a consequence of conflating ``linguistic realism'' \cite{carnap2003} with ``scientific realism'' \cite{arroyodasilva2022prin}.\footnote{Carnap's stance in the realist-antirealist debate should be called ``\textit{frameworkism}'', as he doesn't go beyond his frameworks in any meaningful sense \cite[for a defense of the use of such term]{arroyodasilva2022prin}.}

Which isn't to say that WFR$^{HD}$ has gone instrumentalist all the way. Surely, to go as far as to deny that the world has an objective structure and to affirm that scientific theories are merely instruments \textit{is} to move towards an anti-realist stance. However, as Ladyman already put it, ``[a]n anti-realist about science doesn't have to deny that there's an objective reality; they can just deny that we're finding out about it'' \cite[607]{ladyman2010manyworldstranscript}. Both the realist and the anti-realist stances come in many colors and shapes. Constructive empiricism is a well-known anti-realist stance towards scientific theories which argues that we should not maintain belief in the truth of scientific theories; alternatively, we should adopt a position of epistemic humility and accept scientific theories as empirically adequate \cite{vanfraassen1980}. The constructive-empiricist stance, then, denies that we should reify the items of one's ontology as true as we cannot decide the correct framework on the basis of science alone, the multiple frameworks fleshed out in so many different ways according to which the world could be if each framework were true \cite{vanfraassen1989}.

Ney is emphatic in stating that WFR$^{HD}$, as describing a local and separable ontology for quantum mechanics, comes from ``broadly philosophical considerations'' and ``brute intuition'' \cite[87]{ney2021routledge}. For one thing, as David Wallace puts it, it is unclear how fruitful the appeal to intuitions is, given the question of ``\textelp{} why expect our intuitive faculties, evolved as they are for the emergent classical world, to track truth about fundamental metaphysics?'' \cite[71]{wallace2021routledge}. Now, we stress this because there is a crucial element that seems to be overlooked by WFR$^{HD}$: the relation between the realist part of the enterprise and its relation with truth. Pragmatic and intuitive considerations are not truth-conductive.

Beginning from the latter, it suffices to point out that science can (and arguably often does) reveal a counterintuitive picture of reality, so one can safely say that intuitiveness is not a tool that works well to ground claims for scientific realism. That is, when it comes to considering the epistemology of ontology (the metaontology), intuition is not a guide to the truth of a framework. Going to the former, choosing a theory based on pragmatic components is a characteristic of constructive empiricism. Textbook scientific realism, recall, is characterized as a stance according to which we should believe scientific theories because they are true. In particular, we should believe in the actual existence of its ontological commitments. The anti-realist, as depicted by constructive empiricism, relativizes ``belief'' in terms of ``acceptance'' and ``truth'' in terms of ``empirical adequacy'', viz. it exchanges epistemic aspects by pragmatic reasons: ``\textelp{} the amount of belief involved in acceptance is typically less according to anti-realists, they will tend to make more of the pragmatic aspects'' \cite[13]{vanfraassen1980}. Thus, WFR$^{HD}$ fails to provide support for the realist stance towards the existence of wave functions of interest for the realist (in the robust, metaontological sense): the appeal to pragmatic considerations and intuitiveness are at odds with scientific realism. As Otávio Bueno stresses: ``\textelp{} if realists \textit{only} had pragmatic reasons for the acceptance of theories, there wouldn't be any reason to entitle them to claim that the selected theory is \textit{true} (or approximately so)'' \cite[99, original emphasis]{bueno2018}.\footnote{Remember also that Ladyman and Ross have blamed abuse of intuition as the source for the actual failure of analytic metaphysics as a legitimate epistemic enterprise \cite{ladymanross2007}. According to them, intuition leaves us very far away from the image of reality delivered by actual science.}

This, of course, bears on the standard way of understanding scientific realism and its relationship to pragmatic or theoretical virtues. In this view, scientific realism is not typically grounded in considerations such as simplicity, parsimony, or explanatory power. See, for instance, how Nina Emery draws a clear distinction between pragmatic and truth-conducive support, both in the context of science and of metaphysics (it goes without saying that her ``realism about science'' just is scientific realism). This distinction reinforces the point that, for a view to count as genuinely realist, its justification must ultimately appeal to considerations that track truth---not merely to those that are practically or heuristically useful.

\begin{quote}
\textit{Realism about science.} Our best scientific theories are true theories about what the world is like.\\
\textit{Pragmatism about science.} Our best scientific theories are theories that are useful for creatures like us in navigating the world.
\\\textelp{}\\
\textit{Realism about metaphysics.} The aim of metaphysics is to put forward theories about what the world is like that are true.\\
\textit{Pragmatism about metaphysics.} The aim of metaphysics is to put forward theories about what the world is like that are merely useful for creatures
like us. \cite[54--55, original emphasis]{emery2023}.
\end{quote}

It certainly remains an open question whether these virtues might play a more central role. Attempts to go beyond the default stance on theoretical virtues and scientific realism can be found in, for example, \cite{dawid2013,keas2018,schindler2018}. Still, considerable groundwork would be needed to overturn the standard view. In any case, neither WFR nor scientific realism---as standardly construed---seeks support in purely pragmatic terms. The key issue here is that pragmatic virtues, while useful, are not themselves truth-conducive. As such, appealing to them falls short of justifying a genuinely realist commitment, crucially metaontologically.
\color{black}

\section{Between realism and empiricism}

But this is precisely what is out of hand with WFR$^{HD}$. To be sure, here's what is meant by ``realist'' in the context of WFR$^{HD}$:

\begin{quote}
    By a `realist version of quantum mechanics', I mean one that takes the theory to be aimed at providing a \textit{true description of a world} independent of us as observers. \textelp{} In general, realist versions of quantum mechanics are intended to be \textit{descriptive of an object or objects that exist independently of us} or any other observer. \cite[5--6, emphasis added]{ney2012nous}.
\end{quote}

This is remarkable, as the discourse about the world goes beyond what WFR$^{HD}$ actually provides. If realism requires specifying what \textit{truly happens in the world independently of us}, then mere ontological realism is insufficient. In that case, one must also be a realist at the metaontological level. Yet, the connection between these two levels falls short \textit{if} it is grounded merely in pragmatic criteria. One might argue that pragmatic virtues are being used as a sort of `tie-breaker', but this too fails to suffice if realists are to care about the truth of their ontological frameworks---that is, the metaontological question. To qualify as a realist stance, WFR$^{HD}$ would need to go as far as asserting that there are wave functions \textit{in the world, independently of us}. WFR$^{HD}$ acknowledges this point---see the extended quotation above on what a ``realist version of quantum mechanics'' should aim for---but ultimately does not deliver it. Hence, we ought to temper our expectations. On the metaontological level, WFR$^{HD}$ offers pragmatic reasons for preferring it over alternative views; it does not offer alethic features that truly correspond to the world and thereby compel belief in its ontology. That's what's missing---and that's precisely what one cannot have by playing with Carnapianism in the philosophy of science in general.

One might argue, however, that underdetermination prevents an affirmative answer to the metaontological question.  After all, there are alternatives to WFR (again, \textit{qua} WFR$^{HD}$ or else), just as there are alternatives to pretty much every position in quantum mechanics \cite{callender2020}. This is something contemporary scientific realists are well aware of---hence their embrace of fallibilism. So is Ney \cite{ney2021oup}. Recall that the principle of tolerance encourages a humble attitude toward which framework to adopt. That is, even realists may suspend judgment on the correct framework. Yet this humility is merely provisional for scientific realism. Since the realist is committed to the idea that there \textit{must} be a true description of the world, they are also committed to the claim that there \textit{must} be a true framework---and that the task of science is to find it. In this sense, the permanent adoption of epistemic humility ultimately undermines the realist stance.

Let us press this point a little further. Here's a typical scientific realist statement:

\begin{quote}
    \textelp{} when scientists working with the Large Hadron Collider (LHC) in Geneva announced that they had found the elementary particle known as the Higgs boson in 2012, they were not merely making things up. \textit{Rather, they were talking about a real thing that exists in nature}, even though this real thing, namely, the Higgs boson, is unobservable, that is, it takes sophisticated scientific instruments, such as particle accelerators, colliders, and the like, to detect elementary particles like the Higgs boson. \cite[21, emphasis added]{mizrahi2020}.
\end{quote}

So there are bosons and fermions within the Standard Model of particle physics. This is an ontological statement: there are such and such entities \textit{modulo} such and such theories. This is uncontroversial. But is this realist \textit{enough}? Notice how Moti Mizrahi frames the situation: ``they were talking about a real thing that exists in nature'' \cite[21]{mizrahi2020}. Now, that's more like the scientific realism we all know and love.

Back to WFR$^{HD}$. There are wave functions \textit{modulo} WFR$^{HD}$, and there are pragmatic virtues at stake for us to choose WFR$^{HD}$ over other ontologies of quantum mechanics that populate the theory with different entities. 

Empirical virtues alone cannot adjudicate between equally empirically adequate interpretations or theories. This is precisely why extra-empirical virtues are called upon to play a central role in theory choice in the first place \cite{ruetsche2024}. This is particularly evident in quantum mechanics, where ontological underdetermination between distinct interpretations is a persistent issue \cite{callender2020,fraser-vickers2022}. For example, ontological parsimony---invoked through Ockham's razor---is often raised as an argument against the many-worlds interpretation \cite{wallace2012}. Likewise, the appeal to intuitive plausibility is frequently treated as a guiding principle for theory choice \cite{emery2017}. Ney, for instance, defends her interpretation not only on the grounds of its intuitive appeal---via a separable and local ontology---but also for its simplicity: ``It is just simpler to believe in a basic set of entities each with their own locations upon which everything else depends'' \cite[129]{ney2021oup}.

While the usefulness of such extra-empirical criteria for theory choice is not in question, one may challenge the idea that some of them are truth-conducive. One way to articulate this concern is to note that simplicity and intuitiveness pertain to \textit{us}, not to the world.

Non-empirical virtues do play a role in the \textit{acceptance} of theories---and perhaps this is the best thing we can do at the moment \cite[311]{dacostabueno2011}---but they shouldn't play a role in belief. And if that's the case, then the WFR$^{HD}$ is \textit{accepted} (not believed), which is a distinguishing trait of constructive empiricism and is different from the realist claim of believing in its truth \cite{vanfraassen1980}. ``Belief'', as Paul Teller eloquently puts it, ``unlike acceptance, involves thinking that `this is the way things are' \textelp{}'' \cite[126]{teller2001}. Belief, in this sense, involves belief in the truth of a theory or claim. It should involve alethic claims, not pragmatic. Whereas the constructive empiricist would make a much more epistemically humble claim, namely: ``\textelp{} perhaps the world \textit{could be} that way'' \cite[264, original emphasis]{vanfraassen1991}.

There is of course space for disagreement here, so one could raise the following objection, claiming that this is pretty much standard for every physical theory (every specific interpretation of it). If all ontological/existential claims are always made within a framework, the source of disagreement between realists and anti-realists is the metaontological claim---viz., the veracity of a given framework. Without the metaontological claim, however, the realist position becomes indistinguishable from some anti-realist positions, e.g. constructive empiricism. Of course, one does not need to go as far as radical skepticism, idealism, or instrumentalism to claim a position in the anti-realist camp. Constructive empiricists would be happy to go as far as internal ontology goes: wave functions exist within wave function theories. Now, suppose one claims that this is a realist attitude, as Putnam did.
\begin{quote}
    Admittedly, Putnam's position does boast a rich ontology. Electrons exist every bit as much as chairs and tables do, and electrons can even help to \textit{explain} the superficial properties of macro-objects. Few realists, however, are willing to count this as a sufficient condition for being a ``realist.'' After all, Putnam insists that ontological commitment is always internal to a conceptual scheme; there is no scheme-independent fact of the matter about the ultimate furniture of the universe. \cite[49, original emphasis]{anderson1992}.
\end{quote}
If our best shot is internal realism concerning the ontological aspect of scientific realism; if the best we can do is to state that wave functions exist within wave function theories, then how far is this from saying that every theory is a model of how the world could be? This is not, of course, \textit{all} WFR$^{HD}$ has to say; it is also argued that it is useful to adopt this way of understanding the world, at least for certain purposes \cite[sec.~3.10]{ney2021oup}. Such a humble attitude, however, is equally distant from answering the metaontological question concerning the truth of the framework---or, to put it bluntly, its correspondence with ``the world'', so dear to realism.

\section{Monist empiricism?}\label{sec:5}

To be fair, WFR$^{HD}$ is not centered on the claim that it provides, to use the words of Wallace and Christopher Timpson \cite[221]{wallace-timpson2010}, ``the One True Interpretation'' of quantum mechanics. Way more humbly than that, WFR$^{HD}$ provides an ontological account of quantum mechanics that is worth taking into account, and that should legitimately sit at the table next to other quantum ontologies.

Since it doesn't provide truth-conducing reasons for believing in the objective existence of wave functions, this amount of epistemic humility prevents one from describing how the world is, enabling, at best, to provide an account of how the world can possibly be. Perhaps this is our best shot at the ontology for the unobservable. In this sense, there is not much \textit{realism} left in WFR$^{HD}$, as it does sound much like a constructive-empiricist view. WFR$^{HD}$ is a way the world could be, not how the world is. Thus, Tim Maudlin indeed misses the mark when stating that WFR$^{HD}$ tells us ``\textelp{} the way the world is'' \cite[5]{maudlin2022}.\footnote{Bolder statements on this kind may be found, for instance, in the WFR recently proposed by Albert \cite[20--21]{albert2023}.} The existence of wave functions should be accepted on the grounds of pragmatic virtues. WFR$^{HD}$ doesn't recommend belief in wave functions on the grounds of alethic reasons. This was recently recognized by Ney herself: 
\begin{quote}
    But is my position a realism about the wave function? Well, no. We are overreaching if we take the attitude of belief towards wave functions as fields in a space with the high-dimensional structure of a configuration space. So here is where I land. \cite[14--15]{ney2023}.
\end{quote}

Summing up: in order to be a realist stance, WFR$^{HD}$ should be (metaontologically) intended as a true ontology of the world; pragmatic arguments are offered in favor of the view; no truth-conducing arguments are offered; WFR$^{HD}$ has, strictly speaking, done away with truth. Perhaps, then, WFR$^{HD}$ should be better understood as a form of antirealism, viz. ``wave function pragmatism''\footnote{Not to be confused with the pragmatist approach championed by Richard Healey \cite{healey2012}; ``pragmatism'' in the present context means just that reasons for supporting WFR$^{HD}$ are pragmatic and not---say---alethic.} or ``wave function empiricism''. But then their realist expectations as per ``\textelp{} wave functions are not just real, but are real, objective, physical fields'' \cite[69]{ney2021oup} should be reconsidered and leveled accordingly. As there's a substantial difference between realism and antirealism, this is not a dispute with regards to the view's name, but with consistently aligning its aims with its results.

Bluntly put, because of the way WFR$^{HD}$ is defended, it seems to be \textit{too antirealist} to be a realist view---it's an antirealist position with regard to the metaontological question. Fair enough. But WFR$^{HD}$ has in fact arguments to favor their view over rivals. Unlike die-hard antirealists, WFR$^{HD}$ gives us reasons to lean towards one view over another. The fact that these reasons don't necessarily lead to the truth shouldn't overshadow the fact that WFR$^{HD}$ positions itself as a favored perspective on quantum ontology. Monism, rather than pluralism, is in the background.\footnote{``Monism'' in this context doesn't mean ``wave function monism''. The latter is a term of art referring to the position according to which there exists a single physical object, which is the wave function. By ``monism'', we mean we shouldn't adopt more than one quantum ontology at a time.}

Contrast this with the case of Bas van Fraassen \cite{vanfraassen1972}: he cannot consistently endorse the modal interpretation of quantum mechanics as the true---or most adequate---one \cite{bueno2014}, despite him being the one who proposed it in the first place. This way, WFR$^{HD}$---even qua wave function pragmatism---seems to be out of step with traditional forms of antirealism, e.g., constructive empiricism.

Finally, if that is not enough to put WFR$^{HD}$ in a kind of limbo concerning the realism--antirealism debate, consider again the quote by one of the advocates of WFR$^{HD}$ we have already advanced: 

\begin{quote} 
Of course, we should be realists about quantum theories. Quantum theories, like the quantum field theories that make up the Standard Model of particle physics, have been extremely well-confirmed. And we should be realists about quite a lot of the phenomena quantum theories reveal. \cite[15]{ney2023}
\end{quote} 

\noindent By plugging the requirement that we should be realists about quantum mechanics, without requiring that this realism have any specific ontology (wave function, primitive ontology, minds, and so on), we are adhering to a very obscure kind of realism, one that invites us to believe in the posits of a theory, without determining which posits are those. That is exactly what realists themselves find problematic. Chakravartty \cite[26]{chakravartty2007}, for one, argued that one just cannot be a realist without having a clear picture of what one is being a realist about \cite{french2018}. As a result, the kind of realism that survives after Ney, within WFR$^{HD}$, recognizes the limits of the reasons to go realist on wave functions is exactly one that does not really fulfill the basic requirements to be realism. Given that antirealism does not suit its initial goals, it is difficult to determine what WFR$^{HD}$ really amounts to. 

Of course, not all WFR proposals suffer from such a fate. In comparison with other paradigmatically realist approaches, the antirealism present in WFR$^{HD}$ becomes even clearer.

As a case in point, take the ``wave function space realism (or wave function space fundamentalism)'' from Jill North \cite[200]{north2013}. This is the view according to which the wave function in a high-dimensional space is fundamental because it is required by the dynamical laws of quantum mechanics (e.g. the Schrödinger Equation). The realist qua metaontological questions are always put forth by North. Take, for instance, the following claim:

\begin{quote}
\textelp{} the dynamical laws are a guide to the fundamental nature of a world. \textelp{} The dynamical laws govern the fundamental level of reality; that is why they are a guide to the fundamental nature of a world. \textelp{} When I say that the laws ``are a guide to the fundamental nature of a world,'' I mean that we infer the fundamental nature of a world from the dynamical laws. We do not directly observe the fundamental level of reality: we infer it from the dynamics. We posit, at the fundamental level, whatever the dynamical laws presuppose---whatever there must be in the world for these laws to be true of it. \cite[186]{north2013}
\end{quote}

North does exactly what is expected from scientific realists, which is to go as far as stating that the world truly has such and such features, and that we could learn such and such features (viz., the existence of a high-dimensional wave function) from scientific theories themselves. The ``general principle'' according to which we can read off ontological commitments from the dynamics enables the realist move according to which theories correspond to the state of affairs in the world. Here's North on this:

\begin{quote}
So why not take the wave function and its space as mathematical tools that do not represent physical things in the world? Because of our guiding principle. This principle says to infer, from the mathematical structure needed to formulate the dynamical laws, the corresponding physical structure and ontology to the world. \cite[191]{north2013}
\end{quote}

\begin{quote}
\textelp{} we generally posit, in the physical world, the fundamental structure and ontology presupposed by the dynamical laws. This match between dynamics and world is evidence that this \textit{is} the fundamental nature of a world governed by that dynamics. \cite[201]{north2013}
\end{quote}

As another example of a realist WFR, we might recover Albert \cite{albert2023} and his argument for two kinds of spaces, one high-dimensional of fundamental wave functions and one three-dimensional of ordinary objects. Albert \cite{albert1996} goes on to say that the fundamental space is all what there is, and that our experience of living in a three-dimensional space is false:

\begin{quote}
And whatever impression we have to the contrary (whatever impression we have, say, of living in a three-dimensional space, or in a four-dimensional space-time) is somehow flatly illusory. \cite[277]{albert1996}
\end{quote}

Again, on the methodological level we're pressing, the relevant thing to emphasize is the movement towards truths in the world, which is characteristic of scientific realism---and absent in the case of WFR$^{HD}$.

\section{Concluding remarks}
Alyssa Ney's interpretation of wave function realism (referred to as ``WFR$^{HD}$'') is notably misaligned with realism, as indicated by the manner in which the proposal was defended \cite{ney2021oup,ney2021routledge}. The author recently acknowledged this divergence from the realist standpoint \cite{ney2023}. However, in face of the difficulties we have discussed in our paper, it may be necessary to reconsider the objectives of the proposal, as WFR$^{HD}$ aligns more closely with antirealists' objectives (hence, the label `Wave Function Pragmatism' we have suggested for it). In light of the proposal's monist foundation, it remains to be determined what form of antirealist position Wave Function Pragmatism will incarnate. So far, we don't know where Wave Function Pragmatism sits in the table of epistemic stances; yet we know where it \textit{does not}: with realism.

The fact is that a commitment to describing things \textit{in the world as they truly are} is precisely what one expects from a full-blown realist attitude---whether directed at science in general or at wave functions more specifically---, and that is what is lacking in WFR$^{HD}$, as we have seen. By contrast, advocating a more cautious and humble stance toward the scientific enterprise aligns more closely with empiricism. When defending a philosophical position---such as wave function realism, or more broadly, scientific realism---it may be unadvisable to adopt the argumentative posture of one's opponents. Doing so not only risks attracting the very criticisms originally directed at them but also threatens to undermine the original motivations for advancing the stance in the first place. Scratch the surface, and the wool unravels: what looked like realism is revealed as antirealism in sheep's clothing, precisely because it evades the metaontological commitment to the framework's truth.

\section*{Acknowledgements}
An earlier version of this text was presented at the VIII International Workshop on Quantum Mechanics and Quantum Information: Quantum Theory and Reality (2022). We would like to thank the audience of the workshop for their comments and criticism which contributed to the improvement of the draft. We also thank Alyssa Ney, Laura Ruetsche, and Otávio Bueno for their encouragement, feedback, and helpful comments on earlier versions of the manuscript.

\section*{Funding}
Raoni Arroyo was supported by grants \#315067/2025-0 and \#446478/2024-5, National Council for Scientific and Technological Development (CNPq). The earlier versions of this article were produced during his research visit to the Department of Philosophy, Communication and Performing Arts of the Unviersity of Roma Tre, Rome, Italy, supported by grant \#2022/15992-8, São Paulo Research Foundation (FAPESP), and during his postdoctoral research conducted at the Centre for Logic, Epistemology and the History of Science (CLE), Campinas, Brazil, supported by grant \#2021/11381-1, São Paulo Research Foundation (FAPESP). Jonas R. Becker Arenhart was partially supported by the National Council for Scientific and Technological Development (CNPq).

\section*{Conflict of interest}
None.
\bibliographystyle{spphys}
\bibliography{Foop.bib}

\begin{thebibliography}{10}
\providecommand{\url}[1]{{#1}}
\providecommand{\urlprefix}{URL }
\expandafter\ifx\csname urlstyle\endcsname\relax
  \providecommand{\doi}[1]{DOI \discretionary{}{}{}#1}\else
  \providecommand{\doi}{DOI \discretionary{}{}{}\begingroup \urlstyle{rm}\Url}\fi

\bibitem{maxwell1962}
G.~Maxwell, {Theories, Frameworks, and Ontology}, Philosophy of Science \textbf{69}, 132 (1962)

\bibitem{chen2019}
E.K. Chen, Realism about the wave function, Philosophy Compass \textbf{14}, e12611 (2019)

\bibitem{albert2013}
D.Z. Albert, in \emph{{The wave function: Essays on the metaphysics of quantum mechanics}}, ed. by A.~Ney, D.Z. Albert (Oxford University Press, Oxford, 2013), pp. 52--57

\bibitem{wallace2021routledge}
D.~Wallace, in \emph{Current Controversies in Philosophy of Science}, ed. by S.~Dasgupta, R.~Dotan, B.~Weslake (Routledge, New York, 2021), pp. 63--74

\bibitem{hubert-2022}
M.~Hubert, {Review of Alyssa Ney's The World in the Wave Function}, Philosophy of Science \textbf{89}, 864 (2022)

\bibitem{maudlin2022}
T.~Maudlin, {Speculations in High Dimensions}, Analysis \textbf{82}, 787 (2022)

\bibitem{chakravartty-sep-scientific-realism}
A.~Chakravartty, in \emph{The {Stanford} Encyclopedia of Philosophy}, ed. by E.N. Zalta, {S}ummer 2017 edn. (Metaphysics Research Lab, Stanford University, 2017)

\bibitem{fine1986}
A.~Fine, {Unnatural Attitudes: Realist and Instrumentalist Attachments to Science}, Mind \textbf{95}, 149 (1986)

\bibitem{french2018}
S.~French, in \emph{The Routledge Handbook of Scientific Realism}, ed. by J.~Saatsi (Routledge, New York, 2018), pp. 394--406

\bibitem{kincaid2018}
H.~Kincaid, in \emph{The Routledge Handbook of Scientific Realism}, ed. by J.~Saatsi (Routledge, New York, 2018), pp. 369--379

\bibitem{maudlin2007}
T.~Maudlin, \emph{The metaphysics within physics} (Oxford University Press, Oxford, 2007)

\bibitem{bird2018}
A.~Bird, in \emph{The Routledge Handbook of Scientific Realism}, ed. by J.~Saatsi (Routledge, New York, 2018), pp. 419--433

\bibitem{hoefer2020}
C.~Hoefer, in \emph{Scientific Realism and the Quantum}, ed. by S.~French, J.~Saatsi (Oxford University Press, New York, 2020), pp. 19--34

\bibitem{callender2020}
C.~Callender, in \emph{Scientific Realism and the Quantum}, ed. by S.~French, J.~Saatsi (Oxford University Press, New York, 2020), pp. 55--77

\bibitem{mizrahi2020}
M.~Mizrahi, \emph{The Relativity of Theory: Key Positions and Arguments in the Contemporary Scientific Realism/Antirealism Debate}, \emph{Synthese Library}, vol. 431 (Springer, Cham, 2020)

\bibitem{ladymanross2007}
J.~Ladyman, D.~Ross, \emph{Every Thing Must Go: {M}etaphysics Naturalized} (Oxford University Press, Oxford, 2007)

\bibitem{ruetsche2020}
L.~Ruetsche, in \emph{Scientific Realism and the Quantum}, ed. by S.~French, J.~Saatsi (Oxford University Press, New York, 2020), pp. 293--314

\bibitem{smart1963}
J.J.C. Smart, \emph{Philosophy and Scientific Realism} (Routledge \& Kegan Paul, London, 1963)

\bibitem{boyd1983}
R.N. Boyd, On the current status of the issue of scientific realism, Erkenntnis \textbf{19}, 45 (1983)

\bibitem{devitt1991}
M.~Devitt, \emph{Realism and Truth} (Blackwell, Oxford, 1991)

\bibitem{kukla1998}
A.~Kukla, \emph{Studies in Scientific Realism} (Oxford University Press, Oxford, 1998)

\bibitem{niiniluoto1999}
I.~Niiniluoto, \emph{Critical Scientific Realism} (Oxford University Press, Oxford, 1999)

\bibitem{psillos1999}
S.~Psillos, \emph{Scientific Realism: How Science Tracks Truth} (Routledge, London, 1999)

\bibitem{chakravartty2007}
A.~Chakravartty, \emph{A Metaphysics for Scientific Realism: Knowing the Unobservable} (Cambridge University Press, Cambridge, 2007)

\bibitem{carnap1950}
R.~Carnap, Empiricism, semantics and ontology, Revue International de Philosophie \textbf{4} (1950)

\bibitem{chalmers2009}
D.~Chalmers, in \emph{In Metametaphysics: New Essays on the Foundations of Ontology}, ed. by D.~Chalmers, D.~Manley, R.~Wasserman (Oxford University Press, Oxford, 2009), pp. 77--129

\bibitem{ladyman2010manyworlds}
J.~Ladyman, in \emph{Many worlds? Everett, quantum theory, and reality}, ed. by S.~Saunders, J.~Barrett, A.~Kent, D.~Wallace (Oxford University Press, New York, 2010), pp. 154--160

\bibitem{ney-albert2013wfr}
A.~Ney, D.Z. Albert (eds.), \emph{{The wave function: Essays on the metaphysics of quantum mechanics}} (Oxford University Press, Oxford, 2013)

\bibitem{ney2021oup}
A.~Ney, \emph{The World in the Wave Function} (Oxford University Press, Oxford, 2021)

\bibitem{ney2012philstud}
A.~Ney, Neo-positivist metaphysics, Philosophical studies \textbf{160}, 53 (2012)

\bibitem{maudlin2013}
T.~Maudlin, in \emph{The Wave Function: Essays on the Metaphysics of Quantum Mechanics}, ed. by A.~Ney, D.Z. Albert (Oxford University Press, Oxford, 2013), pp. 126--153

\bibitem{bueno2005}
O.~Bueno, {Dirac and the dispensability of mathematics}, Studies in History and Philosophy of Science \textbf{36}, 465 (2005).
\newblock \doi{10.1016/j.shpsb.2005.03.002}

\bibitem{bokulich2020wfr}
A.~Bokulich, in \emph{Scientific Realism and the Quantum}, ed. by S.~French, J.~Saatsi (Oxford University Press, Oxford, 2020), pp. 185--211

\bibitem{ney2023}
A.~Ney, {Three Arguments for Wave Function Realism}, European Journal for Philosophy of Science \textbf{13}(50) (2023).
\newblock \doi{10.1007/s13194-023-00554-5}

\bibitem{ney2012nous}
A.~Ney, {The Status of Our Ordinary Three Dimensions in a Quantum Universe}, Noûs \textbf{46}, 525 (2012)

\bibitem{howard1989}
D.~Howard, in \emph{{Philosophical Consequences of Quantum Theory: Reflections on Bell's Theorem}}, ed. by J.T. Cushing, E.~M{c}Mullin (University of Notre Dame Press, Notre Dame, 1989), pp. 224--253

\bibitem{teller1986}
P.~Teller, {Relational Holism and Quantum Mechanics}, British Journal for the Philosophy of Science \textbf{37}, 71 (1986)

\bibitem{hubert-romano-2018}
M.~Hubert, D.~Romano, {The Wave-Function Is a Multi-Field}, European Journal for the Philosophy of Science \textbf{8}, 521 (2018)

\bibitem{french2014}
S.~French, \emph{The structure of the world: {M}etaphysics and representation} (Oxford University Press, Oxford, 2014)

\bibitem{frenchladyman2003}
S.~French, J.~Ladyman, {Remodelling Structural Realism: Quantum Physics and the Metaphysics of Structure}, Synthese \textbf{136}, 31 (2003)

\bibitem{allori2013}
V.~Allori, in \emph{{The wave function: Essays on the metaphysics of quantum mechanics}}, ed. by A.~Ney, D.Z. Albert (Oxford University Press, Oxford, 2013), pp. 58--75

\bibitem{wallace-timpson2010}
D.~Wallace, C.~Timpson, {Quantum Mechanics on Spacetime I: Spacetime State Realism}, British Journal for the Philosophy of Science \textbf{61}, 697 (2010)

\bibitem{monton2013}
B.~Monton, in \emph{{The wave function: Essays on the metaphysics of quantum mechanics}}, ed. by A.~Ney, D.Z. Albert (Oxford University Press, Oxford, 2013), pp. 154--167

\bibitem{mizrahi2022}
M.~Mizrahi, {Your Appeals to Intuition Have No Power Here!}, Axiomathes \textbf{32}, 969 (2022)

\bibitem{carnap2003}
R.~Carnap, \emph{The Logical Structure of the World [and] Pseudoproblems in Philosophy}, 2nd edn. (Open Court, 2003)

\bibitem{arroyodasilva2022prin}
R.~Arroyo, G.O. da{\ }Silva, {Taking models seriously and being a linguistic realist}, Principia \textbf{26}, 73 (2022).
\newblock \doi{10.5007/1808-1711.2022.e84309}

\bibitem{ladyman2010manyworldstranscript}
J.~Ladyman, in \emph{Many worlds? Everett, quantum theory, and reality}, ed. by S.~Saunders, J.~Barrett, A.~Kent, D.~Wallace (Oxford University Press, New York, 2010), pp. 597--607

\bibitem{vanfraassen1980}
B.C. van{\ }Fraassen, \emph{{The Scientific Image}} (Oxford University Press, Oxford, 1980)

\bibitem{vanfraassen1989}
B.C. van{\ }Fraassen, \emph{Laws And Symmetry}.
\newblock Clarendon paperbacks (Oxford University Press, Oxford, 1989)

\bibitem{ney2021routledge}
A.~Ney, in \emph{Current Controversies in Philosophy of Science}, ed. by S.~Dasgupta, R.~Dotan, B.~Weslake (Routledge, New York, 2021), pp. 75--89

\bibitem{bueno2018}
O.~Bueno, in \emph{The Routledge Handbook of Scientific Realism}, ed. by J.~Saatsi (Routledge, New York, 2018), pp. 96--108

\bibitem{emery2023}
N.~Emery, \emph{Naturalism Beyond the Limits of Science: How Scientific Methodology Can and Should Shape Philosophical Theorizing} (Oxford University Press, New York, 2023).
\newblock \doi{10.1093/oso/9780197654101.001.0001}

\bibitem{dawid2013}
R.~Dawid, \emph{String Theory and the Scientific Method} (Cambridge University Press, Cambridge, 2013).
\newblock \doi{10.1017/CBO9781139342513}

\bibitem{keas2018}
M.N. Keas, Systematizing theoretical virtues, Synthese \textbf{195}, 2761 (2018).
\newblock \doi{10.1007/s11229-017-1355-6}

\bibitem{schindler2018}
S.~Schindler, \emph{Theoretical Virtues in Science: Uncovering Reality through Theory} (Cambridge University Press, Cambridge, 2018).
\newblock \doi{10.1017/9781108381352}

\bibitem{ruetsche2024}
L.~Ruetsche, The miracles argument meets quantum mechanics: Toward a locavore philosophy of physics, THEORIA. An International Journal for Theory, History and Foundations of Science \textbf{39}(2), 245 (2024).
\newblock \doi{10.1387/theoria.24976}

\bibitem{fraser-vickers2022}
J.D. Fraser, P.~Vickers, {Knowledge of the Quantum Domain: An Overlap Strategy}, The British Journal for the Philosophy of Science  (2022).
\newblock \doi{10.1086/721635}

\bibitem{wallace2012}
D.~Wallace, \emph{{The emergent multiverse: Quantum theory according to the Everett interpretation}} (Oxford University Press, Oxford, 2012)

\bibitem{emery2017}
N.~Emery, {Against Radical Quantum Ontologies}, Philosophy and Phenomenological Research \textbf{95}(3), 564 (2017)

\bibitem{dacostabueno2011}
N.~da{\ }Costa, O.~Bueno, in \emph{Brazilian Studies in Philosophy and History of Science}, \emph{Boston Studies in the Philosophy of Science}, vol. 290, ed. by D.~Krause, A.~Videira (Springer, Dordrecht, 2011), pp. 301--312

\bibitem{teller2001}
P.~Teller, Whither constructive empiricism?, Philosophical Studies \textbf{106}, 123 (2001)

\bibitem{vanfraassen1991}
B.C. van{\ }Fraassen, \emph{{Quantum Mechanics: An Empiricist View}} (Oxford University Press, Oxford, 1980)

\bibitem{anderson1992}
D.L. Anderson, {What Is Realistic about Putnam's Internal Realism?}, Philosophical Topics \textbf{20}, 49 (1992)

\bibitem{albert2023}
D.Z. Albert, \emph{{A guess at the riddle: Essays on the physical underpinnings of quantum mechanics}} (Harvard University Press, 2023)

\bibitem{healey2012}
R.~Healey, Quantum theory: A pragmatist approach, The British Journal for the Philosophy of Science \textbf{63}(4), 729 (2012).
\newblock \doi{10.1093/bjps/axr054}

\bibitem{vanfraassen1972}
B.C. van{\ }Fraassen, in \emph{{Paradigms and Paradoxes: The Philosophical Challenge of the Quantum Domain}}, ed. by R.~Colodny (University of Pittsburgh Press, Pittsburgh, 1972), pp. 303--366

\bibitem{bueno2014}
O.~Bueno, {Constructive Empiricism, Partial Structures and the Modal Interpretation of Quantum Mechanics}, Quanta \textbf{3} (2014).
\newblock \doi{10.12743/quanta.v3i1.19}

\bibitem{north2013}
J.~North, in \emph{The Wave Function: Essays on the Metaphysics of Quantum Mechanics}, ed. by A.~Ney, D.Z. Albert (Oxford University Press, Oxford, 2013), pp. 184--202

\bibitem{albert1996}
D.Z. Albert, in \emph{{In Bohmian Mechanics and Quantum Theory: An Appraisal}}, ed. by J.~Cushing, A.~Fine, S.~Goldstein (Kluwer, Dordrecht, 1996), pp. 277--284

\end{thebibliography}
\end{document}